\documentclass[9pt,twocolumn,twoside]{opticajnl}
\journal{opticajournal} 

\setboolean{shortarticle}{true}

\usepackage{lineno}
\linenumbers 

\title{\HfO{}-based platform for high-index-contrast visible and UV integrated photonics}

\author[1,*]{Oscar Jaramillo}
\author[1]{Vighnesh Natarajan}
\author[1]{Hamim Mahmud Rivy}
\author[1]{Joshua Tensuan}
\author[2]{Leonardo Massai}
\author[1]{Karan K. Mehta}

\newcommand{\AlO}{\ensuremath{\mathrm{Al}_2\mathrm{O}_3}}
\newcommand{\SiO}{\ensuremath{\mathrm{Si}\mathrm{O}_2}}
\newcommand{\HfO}{\ensuremath{\mathrm{Hf}\mathrm{O}_2}}
\newcommand{\ZrO}{\ensuremath{\mathrm{Zr}\mathrm{O}_2}}
\newcommand{\SiN}{\ensuremath{\mathrm{Si}_3\mathrm{N}_4}}

 \nolinenumbers

\affil[1]{School of Electrical and Computer Engineering, Cornell University, Ithaca, NY 14853, USA}
\affil[2]{Institute for Quantum Electronics, ETH Zurich, 8093 Zurich, Switzerland}

\affil[*]{oj43@cornell.edu}

\begin{abstract}
Ultraviolet and visible integrated photonics enable applications in quantum information, sensing, and spectroscopy, among others. Few materials support low-loss photonics into the UV, and the relatively low refractive index of known depositable materials limits achievable functionality. Here we present a high-index integrated photonics platform based on \HfO{} and \AlO{} composites deposited via Atomic Layer Deposition (ALD) with low loss in the visible and near-UV. We show that \AlO{} incorporation dramatically decreases bulk loss compared to pure \HfO{}, consistent with inhibited crystallization due to the admixture of $\AlO{}$. Composites exhibit refractive index $n$ following the average of that of \HfO{} and \AlO{}, weighted by the \HfO{} fractional composition $x$. At $\lambda=375$~nm, composites with $x=0.67$ exhibit $n=2.01$ preserving most of \HfO{}'s significantly higher index, and $3.8(7) $~dB/cm material loss. We further present fully etched and cladded waveguides, grating couplers, and ring resonators, realizing single-mode waveguide loss of $0.25(2)$~dB/cm inferred from resonators of 2.6 million intrinsic quality factor at $\lambda=729$~nm, $2.6(2)$~dB/cm at $\lambda=405$~nm, and $7.7(6)$~dB/cm at $\lambda=375$~nm. We measure the composite's thermo-optic coefficient (TOC) to be $2.44(3) \times 10^{-5}$ RIU/$^\circ$C near $\lambda=397$~nm. This work establishes (\HfO{})$_x$(\AlO{})$_{1-x}$ composites as a platform amenable to integration for low-loss, high-index photonics spanning the UV to NIR. 
\end{abstract}

\setboolean{displaycopyright}{false} 

\begin{document}

\maketitle

Integrated photonics in the visible and UV is essential for a variety of applications \cite{blumenthal2020photonic} spanning bio-chemical spectroscopy \cite{yang2021miniaturization}, neural stimulation and probing \cite{mohanty2020reconfigurable}, and quantum control of trapped-ion \cite{mehta2016integrated,mehta2020integrated, niffenegger2020integrated, kwon2024multi, mordini2024multi, clements2024sub}, neutral atom \cite{christen2022integrated}, and solid-state quantum systems \cite{mouradian2015scalable}. Visible and UV functionality often presents a challenge however \cite{de2021materials, moody20222022} since commonly used integrated photonics materials absorb at short wavelengths, and scattering loss increases with decreasing $\lambda$ \cite{barwicz2005three,lacey1990radiation}, e.g. as $\sim 1/\lambda^4$ for Rayleigh scattering. 

\SiN{} is a mature and commonly used CMOS-compatible platform with refractive index $n\sim 2.1 $ at $\lambda=405$~nm enabling passive and thermo-optic functionality across much of the visible. However, material loss limits performance below $\lambda\sim 450$~nm \cite{corato2024absorption,niffenegger2020integrated}, with SM waveguides exhibiting ${>}10$ dB/cm loss near $\lambda\sim 405$ nm that further rapidly increases with decreasing wavelength  \cite{sorace2019versatile}. Crystalline AlN waveguides on sapphire \cite{lu2018aluminum} offer loss as low as $\sim$8 dB/cm at $\lambda\sim 390$ nm \cite{liu2018ultra}), but more flexibly integrable AlN deposited on amorphous substrates typically exhibits high loss in the visible and UV due to its polycrystalline form, e.g. 650 dB/cm at $\lambda\sim 400$ nm \cite{stegmaier2014aluminum}. Deposited amorphous aluminum oxide (\AlO{}) \cite{west2019low} has enabled single mode (SM) propagation as low as ${\sim}1.35$~dB/cm at $\lambda=369$~nm \cite{kwon2024multi, garcia2024uv}. Despite its utility for low-loss UV photonics, its relatively low index ($n=1.68$ at $\lambda=405$~nm) and hence low index contrast with respect to \SiO{} is a key limitation. High index contrast is critical for strong optical confinement, small bend radii and device footprints, compact and efficient gratings whose strengths scale approximately as $(n_\mathrm{core}^2-n_\mathrm{clad}^2)^2$ \cite{liu2009photonic,joannopoulos2008molding}, photonic bandgap functionality \cite{joannopoulos2008molding}, and efficient acousto-optic interaction (with acousto-optic figure of merit scaling as $n^6$) \cite{yariv1984optical}. A CMOS-compatible, deposited, high-index material transmissive for blue/UV wavelengths would hence alleviate key limitations of \AlO{} for a variety of integrated photonics functionalities. 

A promising candidate is hafnia (\HfO{}), having a bandgap of approximately 5.65 eV \cite{laegu_kang_electrical_2000}, high refractive index ($n=2.09$ at $\lambda=405$~nm), and well studied CMOS compatible deposition via both sputtering and ALD \cite{kuo1992study, gusev2003ultrathin}. However, its propensity to crystallize in films of thickness beyond a few nanometers \cite{aarik2004optical} results in significant optical loss in waveguides. Its use in photonics has so far been limited to optical coatings or metasurfaces with optical interaction lengths of hundreds of nm \cite{torchio2002high,zhang2020low}; significantly lower losses are usually required in waveguide photonics where  centimeter-scale propagation lengths are typical. Here, we show that incorporation of \AlO{} in a primarily \HfO{} layer results in dramatic decrease in bulk material loss, which we attribute to inhibition of crystallization, consistent with similar observations in the context of electronic applications \cite{ho2002suppressed}. This results in an effective composite with index well described by the fractional average composition of the film, preserving most of \HfO{}'s advantage in index despite drastically lower loss. We also present methods to lithographically pattern these films, demonstrating single-mode waveguide transmission of $2.6(2)$~dB/cm at $\lambda=405$~nm and ring resonators with $2.55\times10^6$ intrinsic $Q$ at $\lambda=729$~nm. Our work establishes (\HfO{})$_x$(\AlO{})$_{1-x}$ composites for high-index visible and UV photonics, as well as fabrication methods for low-loss photonic structures. The resulting platform offers low-loss propagation in a single core material over much of the range offered separately by \SiN{} and \AlO{} to date. 

\begin{figure}[t]
    \centering
    \includegraphics[width=0.5\textwidth]{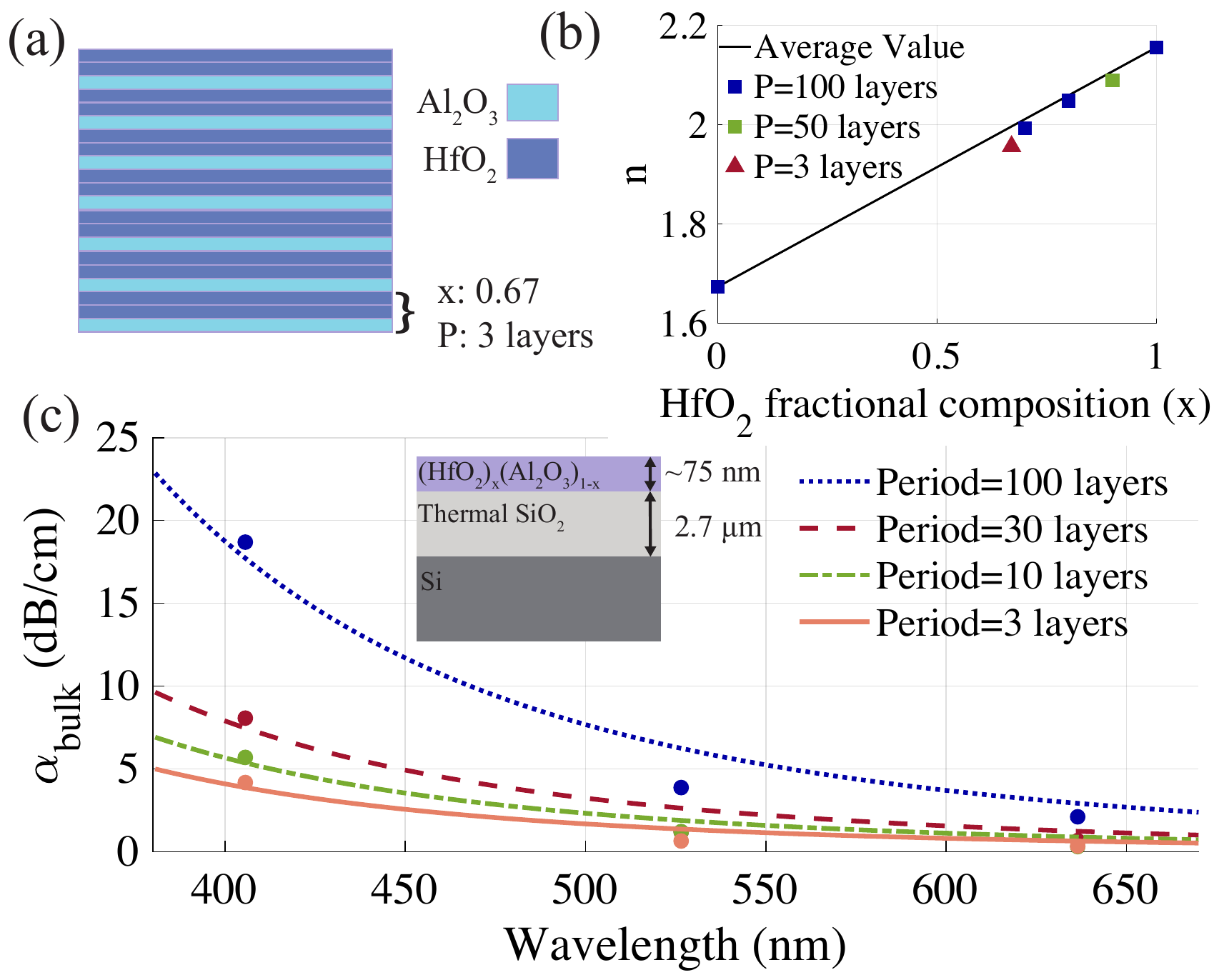}
    \caption{(a) \AlO{} (blue) and \HfO{} (dark blue) layers are deposited with a period $P=3$ and a fractional composition $x=0.67$. (b) Measured refractive index $n$ from an isotropic Cauchy model. (c) Measured material loss ($\alpha_\mathrm{bulk}$) inferred from a slab mode with a fixed fractional composition near $x=0.67$ (data points) with $1/\lambda^4$ fits (lines), corresponding to the assumption of scattering originating from within the bulk. Inset: Film stack-up supporting the slab mode.}
    \label{fig1}
\end{figure}

Composite films were deposited at IBM Zurich using plasma-enhanced ALD (PEALD) at 300$^{\circ}$C with trimethylaluminium (TMA) and tetrakis(ethylmethylamino)hafnium (TEMAH) precursors. Nanolaminate films were grown with layer periodicity $P$, each period consisting of $(1-x) P$ layers of \AlO{} followed by $xP$ layers of \HfO{} (Fig.~\ref{fig1}a shows an example with $x=0.67$ and $P=3$). Films of $\sim$80 nm thickness of varying $x$ were grown on Si wafers, and their $n$ was infered via an isotropic Cauchy fit to spectroscopic ellipsometry data. At $\lambda=405$~nm, the measured $n$ agrees with $xn_{\HfO{}} + (1-x)n_{\AlO{}}$, the weighted average of the two constituent material indices. For propagation loss measurements via a prism coupling method, ${\sim}75$ nm-thick films were grown on Si wafers with 2.7 $\mu$m of thermal \SiO{}, with a fixed composition near $x=0.67$ and varying period from 3 to 100 layers. We calculate the power confinement factor ($C_\mathrm{slab}=P_\mathrm{core}/P_\mathrm{total}$) and estimate the bulk absorption as $\alpha_\mathrm{bulk}=\alpha_\mathrm{slab}/C_\mathrm{slab}$, where $\alpha_\mathrm{slab}$ is the measured slab loss, neglecting the loss arising from surface scattering. As shown in Fig.~\ref{fig1}c, inferred material loss  decreases substantially with decreasing $P$. Atomic force microscopy shows the root mean square roughness decreases from $\sim$2.63 nm for pure \HfO{} to $\sim$0.35 nm for a configuration of $x=0.67$ and $P=3$. Both observations suggest the films are more prone to crystallization as the number of continuous \HfO{} layers increases, consistent with other observations \cite{wei2018growth}. Films with $x=0.67$ and $P=3$ achieve $n=1.96$ and $\alpha_\mathrm{bulk} \approx 4.2$~dB/cm at $\lambda=405$~nm with no annealing. This represents a reduction of $9\%$ in the index of refraction from pure \HfO{} and a substantial decrease in bulk loss from well above the measurement limit of the prism coupling method used (${>}30$~dB/cm) observed for pure \HfO{} at these thicknesses. We refer to this configuration as the composite material for the rest of the article. 

\begin{figure}[b!]
    \centering
    \includegraphics[width=0.5\textwidth]{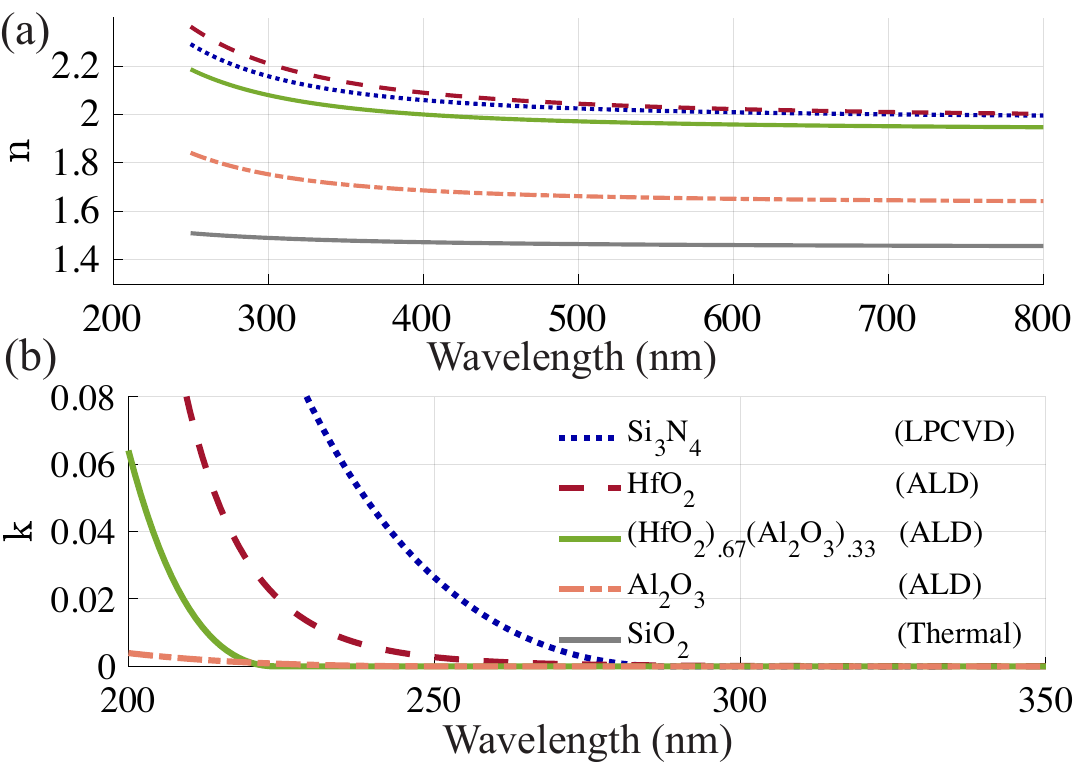}
    \caption{Measured refractive index $n$ (a)  and extinction coefficient (b) of common materials used in photonic devices. Note that $k$ values inferred from ellipsometry indicate onset of strong absorption associated with the optical bandgap, but do not resolve the range of $k\ll0.002$ ($\alpha\ll 1300$~dB/cm at $\lambda=405$~nm) relevant to integrated waveguide devices. } 
    \label{fig2}
\end{figure}

The deposition was reproduced at the Cornell NanoScale Science and Technology Facility (CNF), where after conditioning the PEALD chamber with ${\sim}50$ nm of composite material and annealing in N$_2$ ambient at 800$^\circ$C for one hour, we infer $\alpha_\mathrm{bulk}=2.57(8)$~dB/cm at $\lambda=405$~nm. After annealing, the material is expected to be isotropic, with ALD layering serving only as a means to achieve the mixture. Fig.~\ref{fig2}a shows measured refractive indices for \SiN{}, \AlO{}, \HfO{}, and the composite. We also fit data to a Cody-Lorentz model in order to extract the extinction coefficient. These data demonstrate an improvement in UV transparency (Fig.~\ref{fig2}b) compared to \SiN{} and \HfO{}, with comparable refractive index. To measure the extinction coefficients well below the sensitivity limit of the ellipsometry employed ($k\sim 0.002$, $\alpha\sim 300$~dB/cm) \cite{tompkins2005handbook,corato2024absorption} as relevant to integrated waveguide devices, as well as to explore utilization of this material platform in integrated photonic devices and systems, we fabricate and measure the propagation loss in etched waveguides.

\begin{figure}[t]
    \centering
    \includegraphics[width=0.45\textwidth]{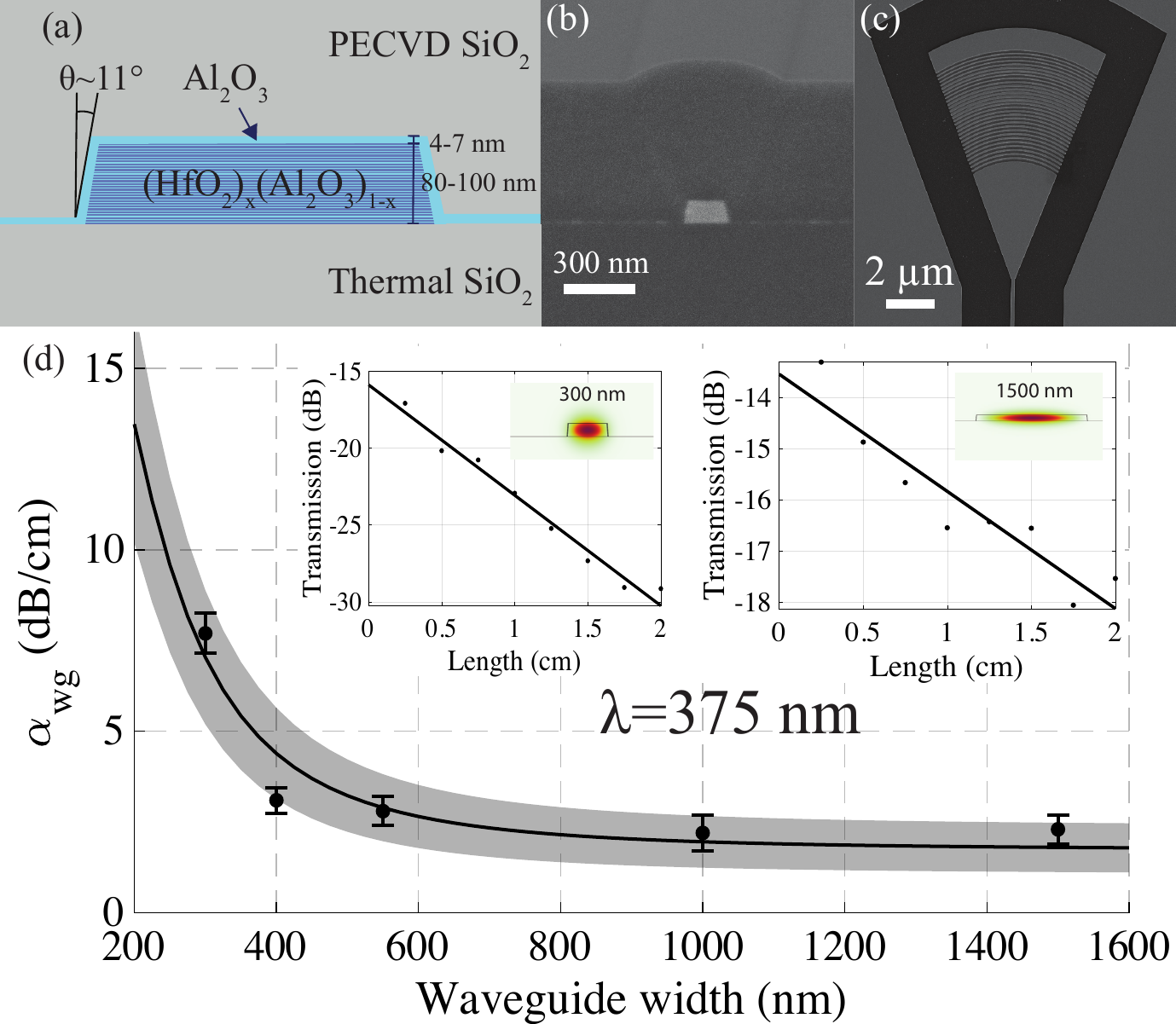}
    \caption{(a) Schematic cross section of composite waveguides realized. (b) SEM cross-section of a 100 nm thick waveguide used to measure waveguide loss at $\lambda=405$~nm. (c)  SEM image of a representative grating used for fiber coupling at $\lambda=405$~nm. (d) Measured waveguide loss at $\lambda=375$~as a function of width for an 80 nm-thick waveguide. Gray line represents the model fit with one standard deviation. Inset: Measured transmission of 300 nm and 1500 nm wide waveguides vs. test length.}
    \label{fig3}
\end{figure}

Device fabrication begins with annealed 80 or 100 nm-thick composite films. Patterns are defined via electron-beam lithography (JEOL9500) in 300 nm of ZEP520-A resist. Patterns are etched into the film via inductively coupled plasma reactive ion etching with a BCl$_3$/Ar chemistry, resulting in a sidewall angle of $\sim$11$^\circ$ (Fig.~\ref{fig3}a). Following a resist strip with remover 1165 and a standard RCA clean, we deposit a $\sim$4 nm layer of \AlO{} and anneal again at $800^\circ$C for one hour. Finally, we deposit $\sim$600 nm of plasma-enhanced chemical vapor deposition (PECVD) \SiO{} at $300^\circ$C using TEOS precursor as a waveguide cladding.

We found that direct deposition of SiO$_2$ with LPCVD at $800^\circ$C or PECVD SiO$_2$ at  $300^\circ$C without the $\sim$4 nm of \AlO{} significantly increased propagation loss as compared to etched air-clad structures. This degradation may be due either to diffusion of oxygen vacancies across the \HfO-\SiO{} interface \cite{capron2007migration}, or the creation of surface states as suspected for \SiN{} \cite{puckett2021422} or Si \cite{borselli2005beyond}. We interpret the thin \AlO{} as acting as a diffusion barrier or passivation layer, with little impact on the optical modes. 

We employ a cutback method to measure waveguide propagation loss $\alpha_\mathrm{wg}$ and power coupling efficiency $\eta$, via fits to $P_\mathrm{out}/P_\mathrm{in}=\eta^2 e^{-\alpha_\mathrm{wg} z}$ with $z$ the straight waveguide length varied up to 2 cm. We use surface grating couplers \cite{taillaert2002out, beck2024grating} for coupling fabricated waveguides to optical fibers angled $25^\circ$ from the surface normal.  Simulations show the grating coupling efficiency $\eta$ for the quasi-TE mode is ${\sim}10\times$ that for the quasi-TM mode; thus, we selectively excite the former by adjusting input polarization to maximize transmission. Measurements at $\lambda=405$~nm employ light from a fiber-coupled laser diode (Thorlabs LP405-SF10). For a SM waveguide of $80\times300$ nm$^2$ cross-section, we obtain $\alpha_\mathrm{wg}=2.6(2)$~dB/cm. The coupling loss is found to be $\eta=10.8(1)$~dB (simulated to be $\sim$ 5.5~dB) at the nominal design angle without experimental angle optimization. At $\lambda=375$~nm (Toptica iBeam smart) we measure $7.7(6)$~dB/cm for the same waveguide geometry, and $\eta=7.9(3)$~dB (simulated to be 4.1~dB).

Both bulk material loss and sidewall surface scattering contribute to $\alpha_\mathrm{wg}$. For large widths, the sidewall mode overlap decreases and $\alpha_\mathrm{wg}$ is dominated by the bulk. We model $\alpha_\mathrm{wg}$ for waveguides of different widths, taking into account the power confinement factor $C_\mathrm{wg}=P_\mathrm{core}/P_\mathrm{total}$, and surface scattering loss arising from the mode's sidewall overlap (supplemental material). We observe that $\alpha_\mathrm{wg}$'s dependence on waveguide width aligns with the predicted trend (Fig.~\ref{fig3}d). We upper-bound the material loss $\alpha_\mathrm{bulk}$ by assuming all loss arises from the bulk  in a 1500 nm-wide waveguide so that $\alpha_\mathrm{bulk}\approx \alpha_\mathrm{wg}/C_\mathrm{wg}$. At $\lambda=375$~nm, we measure $\alpha_\mathrm{wg}= 2.3(4)$~dB/cm, inferring $\alpha_\mathrm{bulk}=3.8(7)$~dB/cm. A similar analysis for $\lambda=405$~nm (supplemental material) gives an upper bound for material loss of $\alpha_\mathrm{bulk}= 3.7(3)$~dB/cm. The similar material losses obtained at $\lambda=375$~nm and $405$ nm suggests these wavelengths are not near the absorption band edge.

\begin{figure}[b!]
    \centering
    \includegraphics[width=0.45\textwidth]{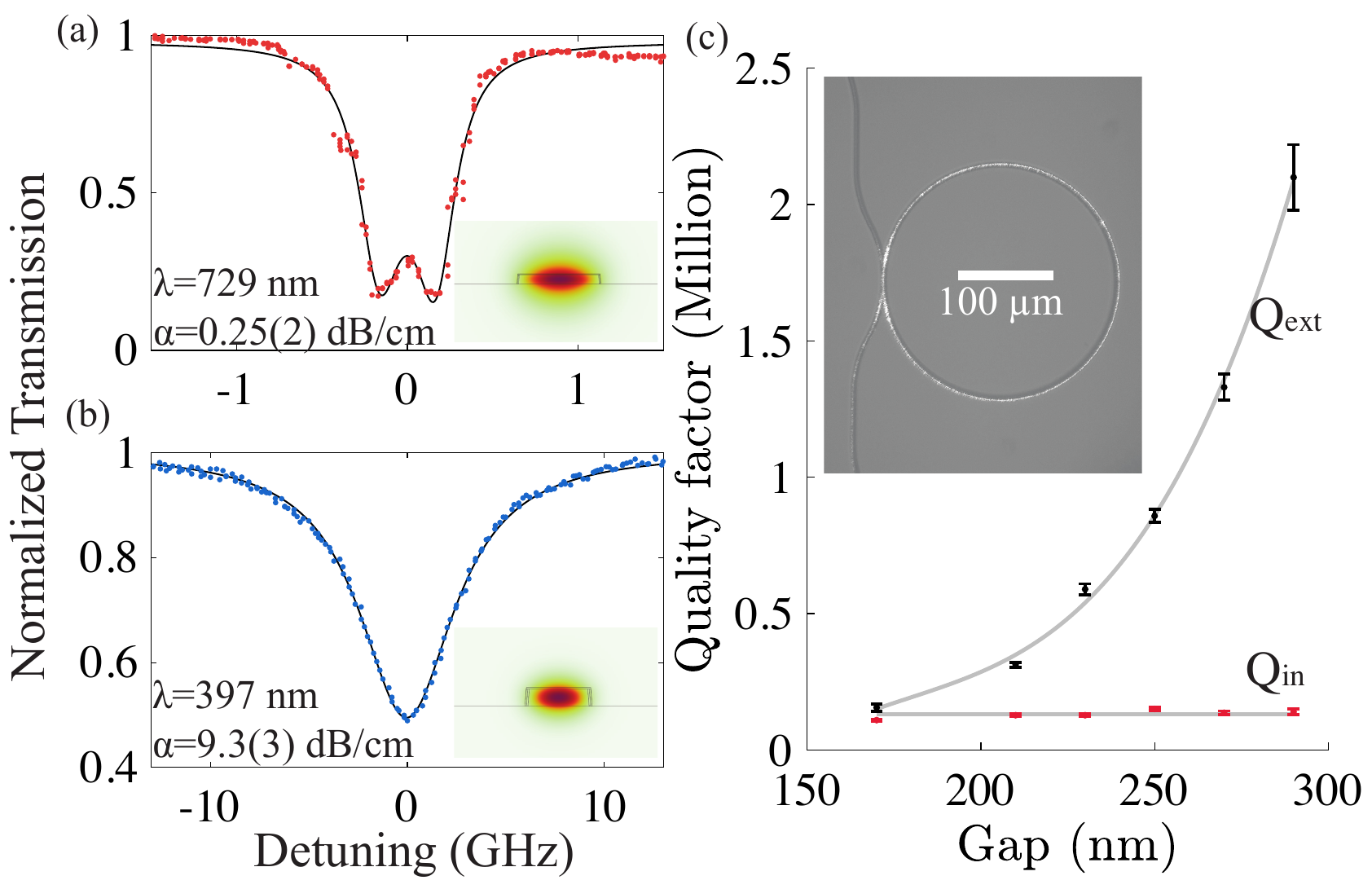}
    \caption{(a) Measured and fit transmission for a nearly critically coupled ring near $\lambda=729$~nm and (b) for $\lambda=397$~nm for a 250 nm coupling gap. (c) Measured $Q_\mathrm{in}$ and $Q_\mathrm{ext}$ at $\lambda=397$~nm. With increasing coupling gap, $Q_\mathrm{in}$ is relatively constant and the $Q_\mathrm{ext}$ increases, indicating reliable extraction of $Q_\mathrm{in}$. Gray lines are polynomial fits to the data as guides to the eye. Inset: Micrograph of ring resonator operating near critical coupling. }
    \label{fig4}
\end{figure}

In the visible/NIR at $\lambda=729$~nm, waveguide loss is low enough to pose a challenge for practical cutback measurement. We fabricate ring resonators in a symmetric coupling configuration with a radius of 250 $\mu$m and waveguides (ring and bus) with a width of $750$ nm ($C_\mathrm{wg}=21\%$), near the threshold of SM operation. Transmission optimization via polarization control through grating coupled waveguide structures ensures quasi-TE mode excitation similar to above. We sweep the frequency of a CW Ti-sapphire laser (M-squared SolsTiS) across the resonance and monitor the frequency with a wavemeter (HighFinesse WS-7). We fit the transmission to obtain the intrinsic ($Q_\mathrm{in}$) and external quality factor ($Q_\mathrm{ext}$). $\alpha_\mathrm{wg}$ is related to $Q_\mathrm{in}$ by 

\begin{equation}
    \alpha_\mathrm{wg}=10 \log_{10}(e)\frac{2\pi n_\mathrm{g}}{\lambda Q_\mathrm{in}} \quad  \text{[dB/cm]},
\end{equation}
where $n_\mathrm{g}$ is the group index of the bend mode and is simulated to be $n_\mathrm{g}=1.683$. A gap of 800 nm results in near critical coupling, exhibiting $Q_\mathrm{loaded} = 1.34\times10^6$ and $Q_\mathrm{in} = 2.55\times10^6$ (Fig.~\ref{fig4}a), corresponding to $\alpha_\mathrm{wg}= 0.25(2)$~dB/cm. We again upper bound the material loss to be $\alpha_\mathrm{bulk}<1.25$~dB/cm near $\lambda=729$~nm.

We also measure ring resonators with a 300 nm width and a 180 $\mu$m radius at $\lambda = 397$ nm, using a tunable external cavity laser (Toptica DL Pro). Devices with varying coupling gaps were measured, showing that the intrinsic quality factor remains constant, while the external quality factor increases, as expected in an under-coupled regime (Fig.~\ref{fig4}c). A group index of $n_\mathrm{g}=2.03$ together with the fit $Q_\mathrm{in} =1.50(5)\times 10^{5}$ gives $\alpha_{\text{wg}} = 9.3(3)$~dB/cm, ${\sim}2-3\times$ higher than expected based on the SM propagation measurements, despite comparable sidewall overlap of the bend mode for this resonator geometry. We believe this discrepancy is due to suboptimal exposure and proximity correction in the bent waveguides and coupling region, resulting in visible excess loss in this region with the present dosage.


We measure the composite's TOC ($\frac{d n_{\text{core}}}{dT}$) by tuning the temperature and monitoring transmission near  $\lambda=397$~nm. The measured temperature-dependence of the resonance wavelength $\lambda_\mathrm{res}$ is related to the material refractive indices comprising the waveguide ($n_\mathrm{core}$ and $n_\mathrm{clad}$) via
\begin{equation}\label{thermo}
    \frac{d\lambda_{\text{res}}}{dT} = \frac{\lambda_{\text{res}}}{n_\mathrm{g}} \left[ \frac{\partial n_{\text{eff}}}{\partial n_{\text{core}}} \frac{d n_{\text{core}}}{dT} + \frac{\partial n_{\text{eff}}}{\partial n_{\text{clad}}} \frac{d n_{\text{clad}}}{dT} \right].
\end{equation}
\begin{figure}[t]
    \centering
\includegraphics[width=0.49\textwidth]{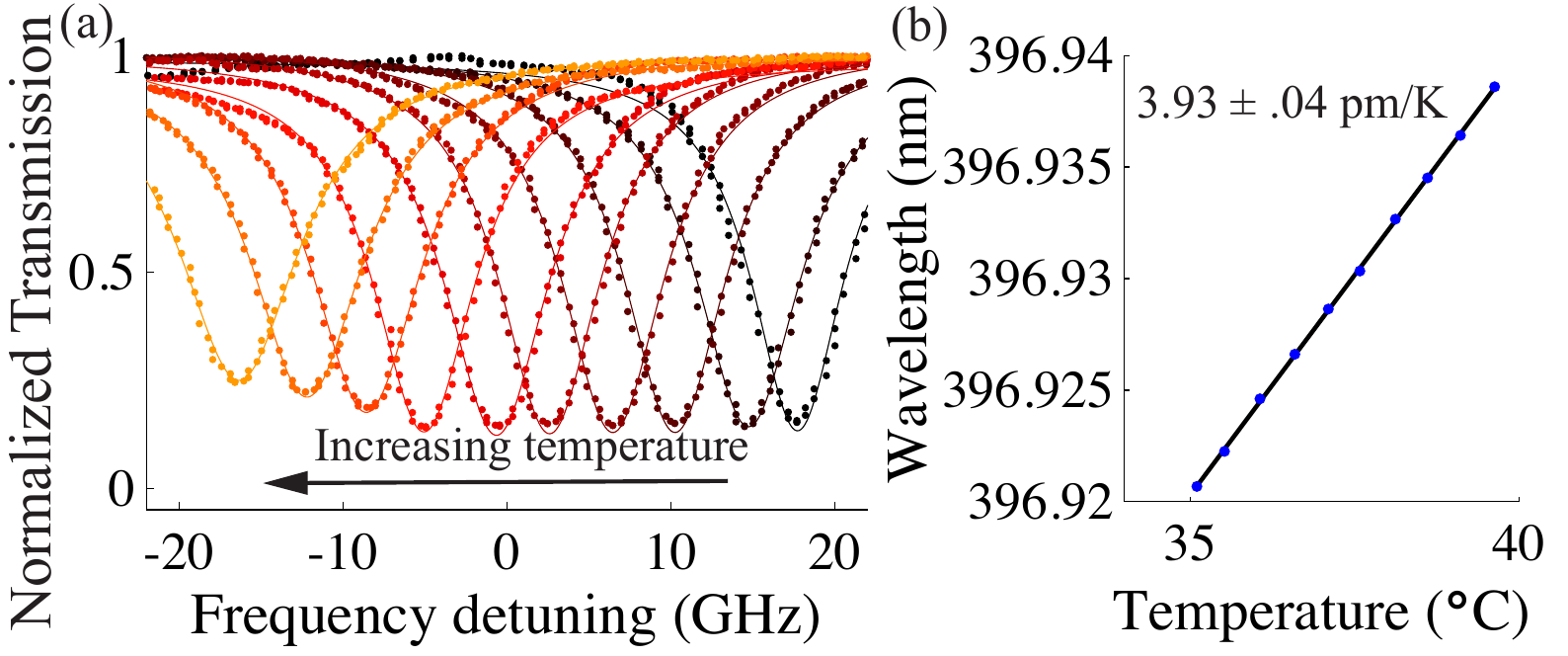}
    \caption{(a) Microring transmission near $\lambda=397$~nm for temperatures between 35.1$^\circ$C to 39.5 $^\circ$C. (b) Resonance wavelength versus temperature. A linear fit shows \mbox{$\frac{d\lambda_{\mathrm{res}}}{dT}=3.93(4)$} pm/K, corresponding to \mbox{$\frac{d n_{\text{core}}}{dT}=2.44(3)\times 10^{-5}$} RIU/K.}
    \label{fig5}
\end{figure}
Fig.~\ref{fig5} shows measured resonance temperature shifts, with fit resulting in $\frac{d\lambda_{\text{res}}}{dT}=3.94(4)$ pm/K. Taking SiO$_2$'s TOC to be $1\times 10^{-5}$ RIU/K \cite{matsuoka1991temperature} and considering the mode's confinement factor we  find $\frac{d n_{\text{core}}}{dT}=2.44(3) \times 10^{-5}$ RIU/K, close to measured values for \AlO{} and \SiN{} \cite{west2019low, elshaari2016thermo}.


These results show that (\HfO{})$_x$(\AlO{})$_{1-x}$ enables low-loss visible and UV photonics, preserving most of \HfO{}'s advantage in refractive index compared to platforms in pure \AlO{}. We have demonstrated fabrication of SM waveguides, microresonators, and grating couplers (see also \cite{smedley2024atomic}). The composite's broad transmission range can enable significant process simplifications, potentially allowing for devices operating over required bandwidths with just one or multiple layers of this composite, rather than both \AlO{} and \SiN{} \cite{sorace2019versatile}. Further work will explore alternatives to PEALD to avoid potential plasma damage \cite{profijt2011plasma}, such as sputtering, which may enable lower material loss \cite{garcia2024uv}, ultimate short-wavelength transmission and power-handling limits for these composites, as well as routes to alleviating surface roughness currently limiting UV SM waveguide loss. Other deposited wide-bandgap materials including \ZrO{} and AlN could be incorporated in similar amorphous composites, for broad transparency and tunable index.

\begin{backmatter}
\bmsection{Funding} 
 We acknowledge support from the National Science Foundation (ECCS- 2301389), Alfred P. Sloan Foundation, Corning Incorporated Foundation, and Cornell University. 

\bmsection{Acknowledgments} 
This work was performed in part at the Cornell NanoScale Facility, an NNCI member supported by NSF Grant NNCI-2025233, and the Binnig and Rohrer Nanotechnology Center at IBM Z\"urich. We thank Yuto Motohashi for contributions to apparatus for resonator measurements, Ronald Grundbacher for assistance with early material depositions at IBM Zurich, and Jonathan Home for support in the early stages of this work at ETH. We thank Jeremy Staffa and Eungkyun Kim for insightful discussions, and Roberto Panepucci, Jeremy Clark, Tom Pennell and Alan Bleier for support in the CNF. 
\bmsection{Disclosures} The authors declare no conflicts of interest.
\bmsection{Data Availability Statement}  Data are available from the authors upon request. 
\bmsection{Supplemental document}
See Supplement 1 for supporting content. 
\end{backmatter}

\vspace{-10pt}


\end{document}